# Simulation of light $C^{4+}$ ion irradiation and its significant enhancement to the critical current density in $BaFe_{1.9}Ni_{0.1}As_2$ single crystals

M. Shahbazi[1], X.L. Wang[1]*, M. Ionescu[2], S. R. Ghorbani[1,3], S. X. Dou[1], K.Y. Choi[4], K.K. Chung[5]

[1]*Institute for Superconducting and Electronic Materials, Faculty of Engineering, Australian Institute for Innovative Materials, University of Wollongong, North Wollongong, NSW 2519, Australia*

[2] *Australian Nuclear Science and Technology Organisation, New Illawarra Road, Lucas Heights, NSW 2234, Australia*

[3]*Department of Physics, Sabzevar Tarbiat Moallem University, P.O. Box 397, Sabzevar, Iran*

[4]*Frontier Physics Research Division and Department of Physics and Astronomy, Seoul National University, Seoul 151-747, Republic of Korea*

[5]*Nano-functional Materials group, Korea Institute of Materials Science, 797 Changwondaero, Changwon, Gyeongnam, 642-831, Korea*

## Abstract

In this work, we report the simulation of $C^{4+}$ irradiation and its significant effects towards the enhancement of the critical current density in $BaFe_{1.9}Ni_{0.1}As_2$ single crystals. $BaFe_{1.9}Ni_{0.1}As_2$ single crystals with and without the C-implantation were characterized by magneto-transport and magnetic measurements up to 13 T over a wide range of temperatures below and above the superconducting critical temperature, $T_c$. It is found that the C-implantation causes little change in $T_c$, but it can greatly enhance the in-field critical current density by a factor of up to 1.5 with enhanced flux jumping at 2 K. Our Monte Carlo simulation results show that all the C ions end up in a well defined layer, causing extended defects and vacancies at the layer, but few defects elsewhere on the implantation paths. This type of defect distribution is distinct from the columnar defects produced by heavy ion implantation. Furthermore, the normal state resistivity is enhanced by the light $C^{4+}$ irradiation, while the upper critical field, $H_{c2}$, the irreversibility field, $H_{irr}$, and $T_c$ were affected very little.

The commercial applicability of Fe-based superconductors[1] relies on their ability to carry high current, which is determined by the effectiveness of pinning sites in these materials. Heavy ion implantation and neutron irradiation are the most effective approaches to introduce effective artificial pinning centres for supercurrent enhancement in both conventional and high temperature superconductors [1-3]. The vortex pinning in high temperature superconductors is controlled by dynamic and static disorder [4]. Dynamic disorder is caused by large thermal fluctuations, and static disorder is caused by material disorder such as twin boundaries and columnar defects, which can be engineered with heavy ion and neutron irradiation. Another way to increase the scattering rate is to introduce defects by ion irradiation, which introduces efficient pinning centres and also enhances scattering significantly. It is well known that columnar defects created by heavy ion irradiation are the most effective pinning sites to pin two-dimensional (2D) pancake vortices in highly anisotropic high temperature cuprate superconductors. Fe-based superconductors have revealed much smaller anisotropy ($\gamma$ = 1-8 at $T \approx T_c$) [5, 6], especially in doped $BaFe_2As_2$ (122) superconductors with $\gamma \approx$ 1-3. Very strong intrinsic pinning strength has been observed in K doped 122 single crystals with rigid vortices, mainly due to small anisotropy [7]. As a result, the point defects induced by neutron irradiation are effective for pinning vortices and enhancing the critical current density, $J_c$, by a factor of 1.5-3 [8]. Heavy ion implantation using ions such as Au [9], Pb [2, 10], and Ta [11] increases $J_c$ by a factor of 3-10 due to the formation of columnar defects [9, 11]. Both neutron and heavy ion irradiation are expensive procedures compared to light ion implantation for large-scale applications. Here, we report the first efforts to create defects by light-ion, $C^{4+}$, irradiation into optimally Ni doped $BaFe_2As_2$ single crystals. The influence of carbon implantation on the physical properties of $BaFe_{1.9}Ni_{0.1}As_2$ single crystal has been investigated. Our results show that light carbon ion implantation is an effective approach that can significantly enhance in-field $J_c$ with little change in $T_c$. Furthermore, the Monte Carlo simulation indicates that the C implantation only cause distortions to the 122 lattice at a well defined layer, causing little change to the lattice along its implantation paths.

Single crystals with the nominal composition $BaFe_{1.9}Ni_{0.1}As_2$ were prepared by a self-flux method [12]. The as-grown single crystals were cleaved and shaped into thin plates for measurements. Irradiation with 35.59 MeV $C^{4+}$ was carried out perpendicular to the broad surface of the sample, using a square shaped beam 7×7 mm² in cross-section, for a total irradiation time of 3 min with ion flux of $2.5 \times 10^{11}$ ions·cm$^{-2}$. The sample was placed on a conductive sample holder with conductive C-tape, in order to prevent charging and excessive heating during irradiation. The beam current was measured before and after irradiation with a Faraday cup, and the average beam current was approximately 10 nA. The Monte Carlo calculation was used to estimate the distribution of carbon ions and the redistribution of other ions caused by carbon ion collisions. Magnetization was measured using a magnetic properties measurement system (MPMS, Quantum Design). The critical current density was calculated using the Bean model. The transport properties were measured over a wide range of temperature and magnetic fields up to 13 T with applied current of 5 mA using a physical properties measurement system (PPMS, Quantum Design).

Figure 1 shows the distribution of carbon ions in the BaFe$_{1.9}$Ni$_{0.1}$As$_2$ single crystal using the Monte Carlo calculation. The results show that almost all the C ions end up in a well defined layer, at a depth of around 24 μm. This layer looks quite homogenous for 500 carbon ions fired along the red arrow. As the beam of carbon ions is uniformly distributed across the sample surface, we expect a fairly homogenous distribution of carbon in this layer.

The binding energy of BaFe$_{1.9}$Ni$_{0.1}$As$_2$ is about 3 eV/atom, so most of the damage is done by primary carbon ions through primary knock-on collisions and none by the Ba, Fe, Ni, and As recoils, because their energy is below 3 eV, as shown in Fig. 2(a). The energy carried by the C ions into the implanted layer is distributed to the BaFe$_{1.9}$Ni$_{0.1}$As$_2$ crystal lattice, and as a result, the atoms in that layer will recoil or be moved out of their lattice sites. Some of these atoms will fall back into a thermodynamic equilibrium position (self-annealing), but a number of them will remain in interstitial positions, destroying locally the BaFe$_{1.9}$Ni$_{0.1}$As$_2$ lattice. To see which of the lattice atoms are more disrupted by the carbon ions, the calculated distributions of the individual atoms (Ba, Fe, Ni, As) which are knocked out of their lattice sites are shown in Fig. 2(b).

Fig. 2(b) shows that most of the BaFe$_{1.9}$Ni$_{0.1}$As$_2$ lattice disruption is contained in and around the C-implanted layer, at a depth of around 24 μm, with little disruption between the entry surface and the damaged layer. Also, the most disrupted (recoiled) are Fe and As, due to their having the highest concentrations and lower masses. The total number of vacancies produced by C-ions and the Ba, Fe, Ni and As recoils is around 2,300 vacancies/ion in the damaged layer. According to this calculation, the C-irradiation and the resulting C-implanted layer constitute a three-dimensional (3D) defect layer with a thickness of 1.5 µm at a depth of about 24 µm under the irradiated surface. The distribution of damage in the cross-section of this 3D layer has a Gaussian profile. This damage matrix is likely to form a network (connected regions) in the damaged layer. Therefore, the defect/vacancy region coexists with the superconducting region which was not destroyed during C-implantation. This type of defect distribution, which is very similar to extended defects, is distinct from the columnar defects caused by heavy ion implantation [9].

The temperature dependence of the resistivity at zero magnetic field for the sample before and after C-implantation is shown in Fig. 3. The resistivity decreases with decreasing temperature from 300 to 20 K for both the implanted and the un-implanted sample, supporting metallic behaviour of this sample. The resistivity increases from $14.3 \times 10^{-5}$ Ω·cm to $31 \times 10^{-5}$ Ω·cm after carbon implantation at 200 K, which is related to enhancement of impurity scattering after C-implantation. The reduction of $T_c$ after ion implantation is a common feature observed in many high-$T_c$ and pnictid superconductors[3, 8], since it can be affected by different effects such as inter-band scattering [13], a reduction in anisotropy [14], etc. However, the $C^{4+}$ implantation only causes small changes in the $T_c$ and transition width in our sample. The $T_c$ was 18.3 K with a small transition width ($\Delta T_c$) of 0.7 K for the sample without implantation. It decreased very little to 17.8 K with almost the same $\Delta T_c$ (0.8 K) after C-implantation at zero field (Fig. 3). The residual resistivity ratio, $RRR = \rho_n (300 K)/ \rho_n (20 K)$, where $\rho_n$ is the normal state resistivity, decreased from 1.97 to 1.88, indicating enhanced scattering centres after C-implantation.

Enhancement of vortex pinning by the light $C^{4+}$ ion implantation can be clearly seen from the magnetization measurements. Fig. 4(b) shows the magnetization curve at 2 K for the un-implanted and implanted samples. The magnetization in the implanted sample is obviously enhanced. $J_c$ was calculated from magnetic hysteresis data using an extended Bean model: $20\Delta m/ (a (1-a/3b))$ $(a < b)$, where $\Delta m$ is the height of the magnetization loop, and $a$ and $b$ are the length and width of the sample perpendicular to the applied magnetic field, respectively. Fig. 4(a) shows the calculated $J_c$ for pristine and carbon implanted single crystals as a function of field with $H//c$. The implanted sample shows a clearly enhanced $J_c$, which is both field and temperature dependent. At $T = 10$ K, the $J_c$ is enhanced for $H < 4$ T. For $T = 2$ and 5 K, the $J_c$ enhancement persists in both low and high fields. The enhancement ($J_{c\text{-im}}/J_{c\text{-unim}}$) is between about 1.5 and 1 at magnetic field smaller than 4 T, as shown in Fig.4(c).

The peak effect, which has been commonly observed in the Fe-based superconductors, was observed for both implanted and un-implanted samples (Fig. 4). The peak position shifts to lower magnetic field after C-ion implantation, as indicated by the arrows in Fig.4 (a). $J_c$ is as high as $1.6 \times 10^5$ A/cm$^2$ at 5 K at $H = 0.5$ T before C-implantation. The $J_c$ increases to $2.3 \times 10^5$ A/cm$^2$ after carbon implantation. It has been reported that for BaFe$_{1.8}$Co$_{0.2}$As$_2$ crystals irradiated by neutrons with a fluence of $4 \times 10^{17}$ cm$^{-2}$ [8], the $J_c$ increased from $3 \times 10^5$ to $7 \times 10^5$ A/cm$^2$ at $H = 0.5$ T. These results are comparable with those for our C-implanted sample using much lower ion doses of C$^{4+}$ ($10^{11}$ /cm$^2$). Therefore, light C ion implantation could be a very effective and less expensive approach for enhancing the $J_c$ field performance in the Fe-based superconductors for practical applications.

Another feature of BaFe$_{1.9}$Ni$_{0.1}$As$_2$ single crystal is that the pristine and carbon implanted single crystals show the flux jump effect, which is more pronounced in the implanted sample (Fig. 4(b)) as shown in the magnetization hysteresis loops at very low magnetic field and 2 K. The size of the flux jumps is smaller than that observed in Ba$_{0.72}$K$_{0.28}$Fe$_2$As$_2$ single crystal [7, 15], with the flux lines fully penetrating into the whole sample.

In order to further look into the effect of the C ion implantation on other pinning related parameters such as the upper critical field, $H_{c2}$, the irreversibility field, $H_{irr}$, and the pinning potential, we have carried out $R$-$T$ measurements in fields up to 13 T with $H//c$ or $H//ab$. Figure 5 shows the $R$-$T$ curves for the BaFe$_{1.9}$Ni$_{0.1}$As$_2$ single crystal before and after carbon implantation with $H//ab$. The $T_c$ onset slowly shifts to lower temperatures with increasing magnetic field, which is related to the nearly isotropic superconductivity in the 122 family at low temperatures [5]. $H_{c2}$ is estimated as the field at which the resistivity becomes 90% of the normal state resistivity; while $H_{irr}$ is defined by 10% of the normal state resistivity. The $H_{c2}$ in the *ab* plane and along the *c* direction is plotted as a function of temperature in Figure 6. The estimated slopes are -6.65 and -5.28 T/K for $H_{c2}$ and $H_{irr}$ before carbon implantation, and they decline to -6.52 and -4.64 T/K after C-implantation in *H//ab*, respectively. The slopes of $H_{c2}$ and $H_{irr}$ were 2.82 and 2.49 T/K for *H//c* before implantation, and they change slightly to 2.9 and 2.03 T/K after implantation, respectively. It should be noted that the $H_{c2}//c$ was only very slightly enhanced by the C implantation. However, the other parameters were obviously reduced. This is related to the reduction of the electron mean free path due to increasing impurity scattering after C-implantation.

Thermally activated flux flow (TAFF) is responsible for the broadening of the resistivity transition and can be expressed by the following equation: $\rho(T, H) = \rho_n \exp(U_0(T, H)/k_B T)$, where $\rho_n$ is the normal state resistivity, $k_B$ is Boltzmann constant, and $U_0$ is the activation energy. The best fit to the experimental data yields a value of the pinning potential ($U_0/k_B$) of 4100 K at $H < 1$ T for both implanted and un-implanted samples. The $U_0/k_B$ values are shown in Fig. 7. For comparison, we also include $U_0/k_B$ values for $Ba_{0.72}K_{0.28}Fe_2As_2$ [7] single crystals. It can be seen that the $U_0/k_B$ for $BaFe_{1.9}Ni_{0.1}As_2$ single crystal is lower than the reported value of 9100 K for $Ba_{0.72}K_{0.28}Fe_2As_2$ single crystal for *H//ab* [7]. For both $BaFe_{1.9}Ni_{0.1}As_2$ single crystals, the activation energy drops very slowly with increasing applied magnetic field for $H < 1$ T. It can be scaled as $H^{-0.02}$, and then decreases slowly as $H^{0.9}$ for $H > 1$ T for *H//ab*, This is in great contrast to the nearly field independent $U_0$ in $Ba_{0.72}K_{0.28}Fe_2As_2$ single crystals[7], indicating different pinning mechanisms in the Ni and K doped 122 single crystals in high fields.

It should be pointed out that $U_0$ is reduced for both *H//c* and *H//ab* in high fields after C implantation. This means that the pinning strength in the ion implanted sample, which only reflects the pinning energy for fields close to $H_{irr}$ and temperatures close to $T_c$, is weaker compared to the sample without implantation. The observation of reduced $U_0$ at high field can well account for the fact that the C ion implantation does little to change $T_c$, $H_{c2}$, or $H_{irr}$, however, it can enhance the in-field $J_c$ significantly for $H < H_{irr}$. Further investigation on the $J_c$ enhancement is underway using high C ion doses and different energies that can increase both defect density and create extended defects at various implantation depths in the 122 superconductors.

In conclusion, we investigated the effects of C-implantation in $BaFe_{1.9}Ni_{0.1}As_2$ single crystal. Monte Carlo calculation shows that the C ions end up in a well defined layer at a certain depth, causing extended defects and vacancies within the layer, but few defects elsewhere on their paths. It is found that the C-implantation causes little change in $T_c$, but it can greatly enhance in-field critical current density by a factor of up to 1.5, with enhanced flux jumping at 2 K. Our results suggest that light C ion implantation is an effective and cheaper method for the enhancement of $J_c$ in Fe-superconductors compared to the heavy ion implantation and neutron irradiation.

X.L. Wang thanks the Australia Research Council for providing funding support for this work through an ARC Discovery project (DP1094073).

*Email: xiaolin@uow.edu.au

**References**

[1] Y. Kamihara, T. Watanabe, M. Hirano, H. Hosona, J.Am. Chem. Society. 130, 3296 (2008); L. M. Paulius, J. A. Fendrich, W. K. Kwok, A. E. Koshelev, V. M. Vinokur, G. W. Crabtree and B. G. Glagola, Physical Review B **56,** 913 (1997).

[2] R. Prozorov, M. A. Tanatar, B. Roy, N. Ni, S. L. Bud'ko, P. C. Canfield, J. Hua, U. Welp and W. K. Kwok, Physical Review B **81,** 094509.

[3] M. Eisterer, M. Zehetmayer, H. W. Weber, J. Jiang, J. D. Weiss, A. Yamamoto and E. E. Hellstrom, Superconductor Science and Technology **23,** 054006 (2010).

[4] D. R. Nelson and H. S. Seung, Physical Review B **39,** 9153 (1989).

[5] H. Q. Yuan, J. Singleton, F. F. Balakirev, S. A. Baily, G. F. Chen, J. L. Luo and N. L. Wang, nauture **457,** 565 (2009).

[6] S. Weyeneth, R. Puzniak, U. Mosele, N. Zhigadlo, S. Katrych, Z. Bukowski, J. Karpinski, S. Kohout, J. Roos and H. Keller, Journal of Superconductivity and Novel Magnetism **22,** 325-329 (2009).

[7] X.-L. Wang, S. R. Ghorbani, S.-I. Lee,et.al , Physical Review B **82,** 024525 (2010).

[8] M. Eisterer, M. Zehetmayer, H. W. Weber, J. Jiang, J. D. Weiss, A. Yamamoto and E. E. Hellstrom, Superconductor Science and Technology **22,** 095011 (2009).

[9] Y. Nakajima, Y. Tsuchiya, T. Taen, T. Tamegai, S. Okayasu and M. Sasase, Physical Review B **80,** 012510 (2009).

[10] H. Kim, R. T. Gordon, M. A. Tanatar, J. Hua, U. Welp, W. K. Kwok, N. Ni, S. L. Bud'ko, P. C. Canfield, A. B. Vorontsov and R. Prozorov, Physical Review B **82,** 060518.

[11] J. D. Moore and et al., Superconductor Science and Technology **22,** 125023 (2009).

[12] A. S. Sefat, R. Jin, M. A. McGuire, B. C. Sales, D. J. Singh and D. Mandrus, Physical Review Letters **101,** 117004 (2008).

[13] M. Putti, P. Brotto, M. Monni, E. G. d'Agliano, A. Sanna and S. Massidda, EPL (Europhysics Letters) **77,** 57005 (2007).

[14] A. J. Millis, S. Sachdev and C. M. Varma, Physical Review B **37,** 4975 (1988).

[15]K.-Y. Choi, G. S. Jeon, X. F. Wang, X. H. Chen, X.-L. Wang, M.-H. Jung, S.-I. Lee and G. Park, Appl. Phys. Lett., 98, 182505 (2011).

**Figure captions:**

Fig. 1. Carbon ion distribution in the sample after C-implantation.

Fig. 2 (a) The energy of the carbon ions is distributed to the atoms/ions in their paths through collisions. (b) The calculated distribution of the individual Ba, Fe, Ni, and As atoms which are knocked out of their lattice sites.

Fig. 3. Temperature dependence of resistivity for zero magnetic field. The inset shows an enlargement of the transition region.

Fig. 4. The magnetic field dependence of critical current density at different temperatures for implanted and un-implanted samples: (a) ratio of critical current before and after C implantation as a function of temperature at various applied fields; (b) field dependence of magnetization at 2 K for implanted and un-implanted samples; (c) ratio of $J_c$ values with and without ion implantation as a function of temperature at different applied fields.

Fig. 5: Temperature dependence of resistivity for different magnetic fields with field parallel to the *ab*-plane before (right) and after (left) carbon implantation.

Fig. 6. Temperature dependence of the upper critical field before and after carbon implantation for $BaFe_{2-x}Ni_xAs_2$ single crystal.

Fig. 7. Magnetic field dependence of pinning potential for $BaFe_{2-x}Ni_xAs_2$ single crystal before and after carbon implantation. Data for $Ba_{0.72}K_{0.28}Fe_2As_2$ single crystal were taken from Ref. [7].

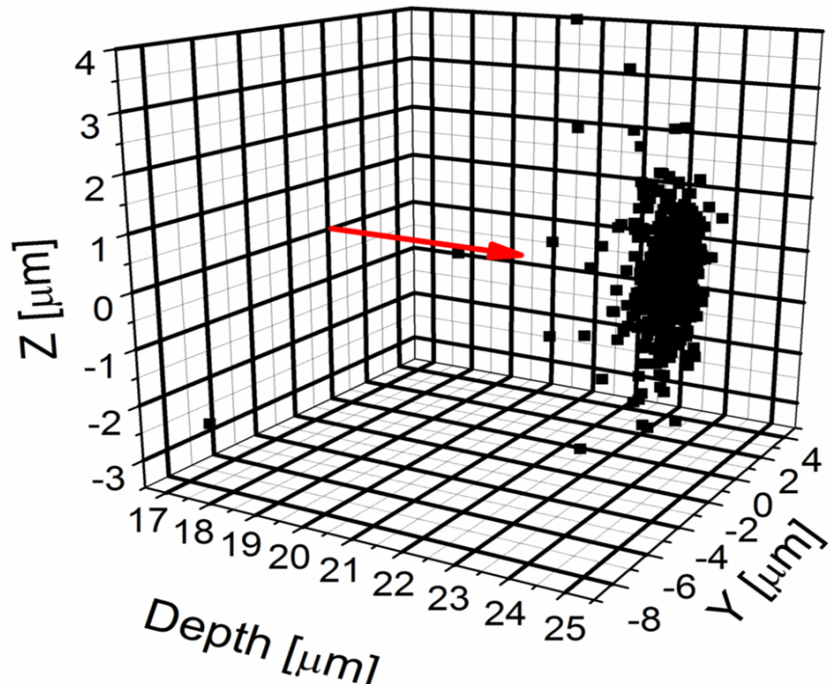

Figure 1

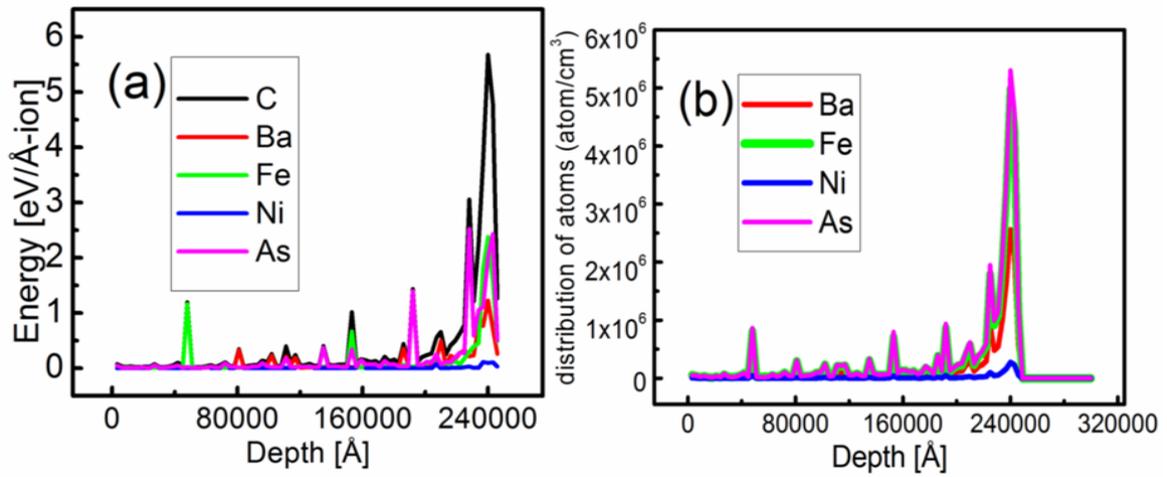

Figure 2

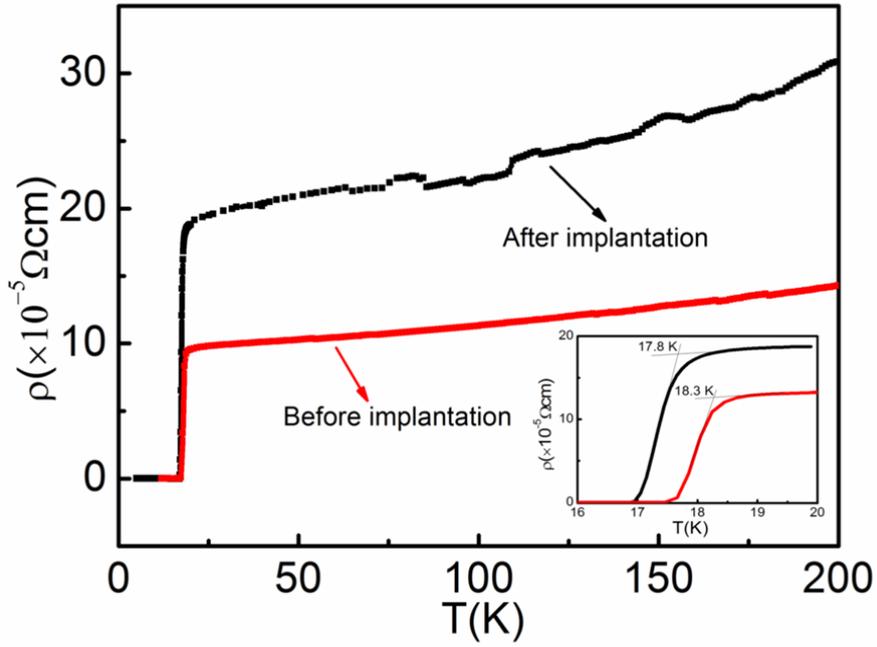

Figure 3

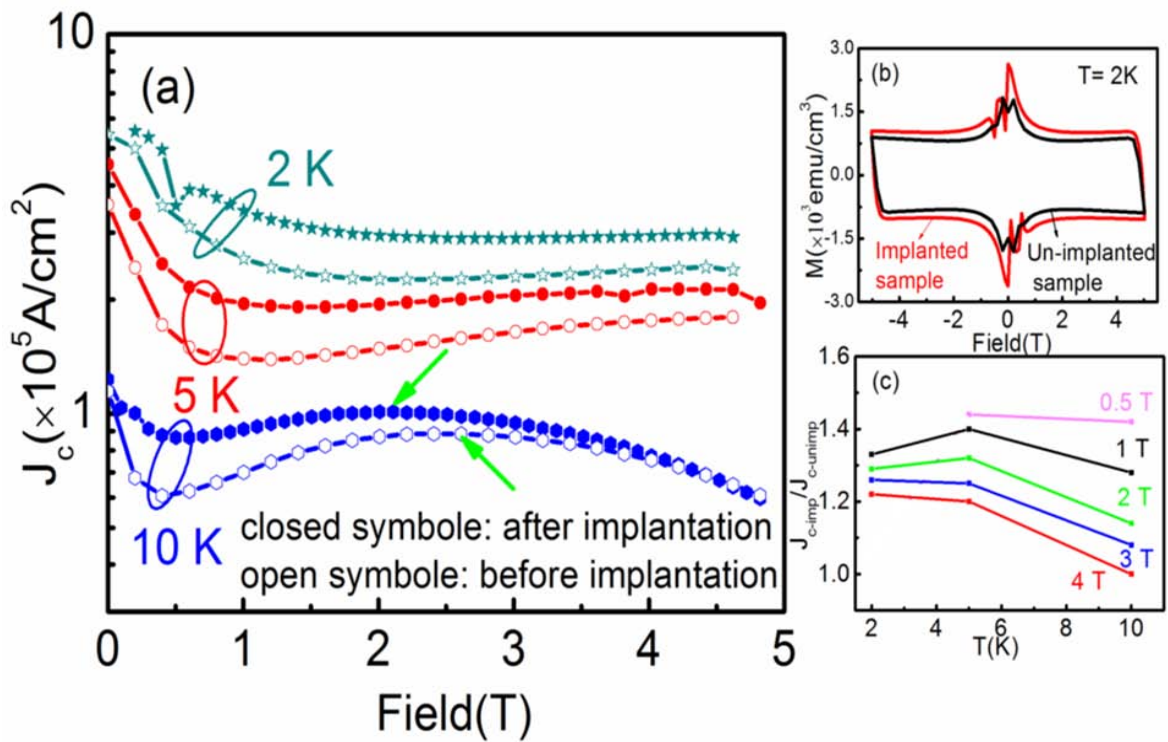

Figure 4

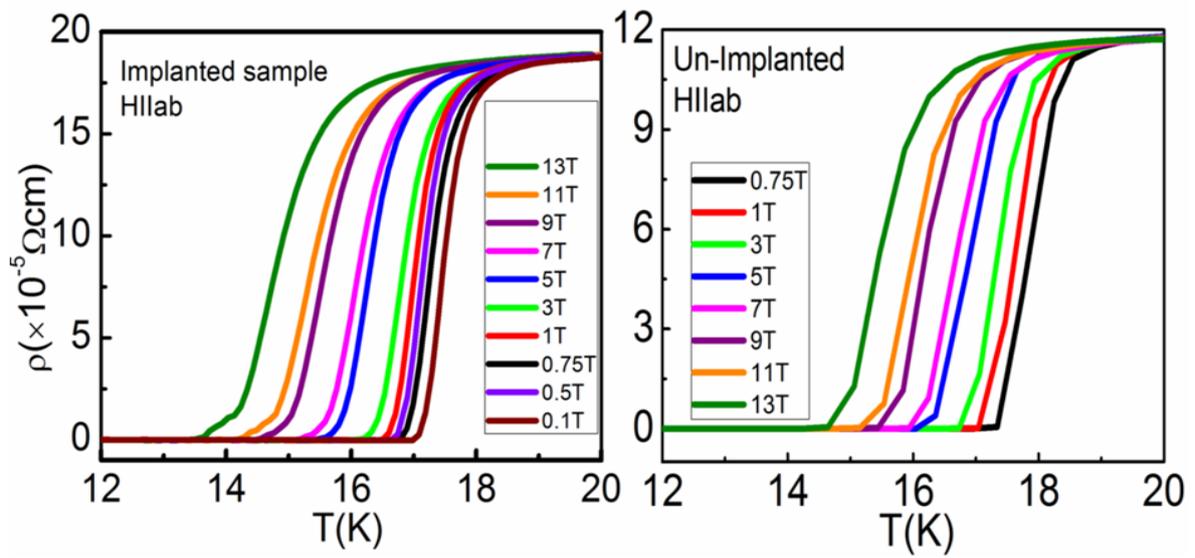

Figure 5

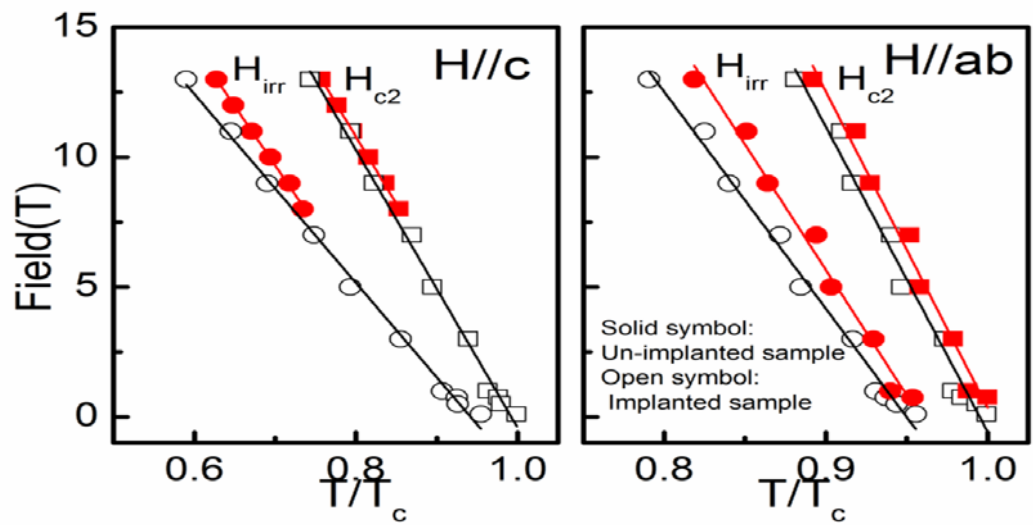

Figure 6

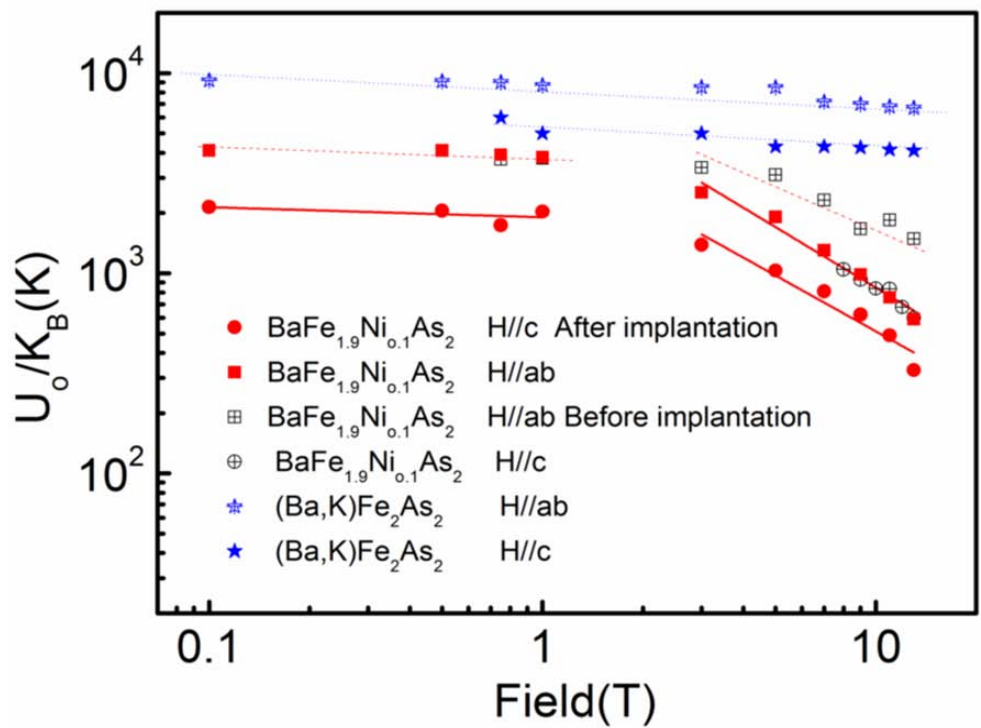

Figure 7